# Sensing the Breath: A Multimodal Singing Tutoring Interface with Breath Guidance

**Ziyue Piao  Gus Xia**

**ABSTRACT**

Breath is a significant component in singing performance, which is still underresearched in most singing-related music interfaces. In this paper, we present a multimodal system that detects the learner's singing pitch and breathing states and provides real-time visual tutoring feedback. Specifically, the breath detector is a wearable belt with pressure sensors and flexible fabric. It monitors real-time body movement of the abdomen, back waist, and twin ribs. A breath visualization algorithm is developed to display real-time breath states, together with the singing pitch contours on an interactive score interface. User studies show that our system can help users not only gain deeper breath during singing but also improve pitch accuracy in vocal training, especially for those with some musical background.

## Author Keywords

*Multimodal Interface, Music Tutoring, Breath Guidance*

## CCS Concepts

•*Applied computing → Computer-assisted instruction; Sound and music computing;*
•*Human-centered computing → Haptic devices;*

## Introduction

Learning to play an instrument is an intrinsically multimodal process, and research has demonstrated that haptic and visual interfaces can help with music education by improving rhythm accuracy [1], playing posture [2], and overall learning efficiency [3]. Unlike playing most musical instruments that focus on finger movements, learning to sing involves the control of the vocal cords and the whole "body instrument". Because inner body movements cannot be seen immediately and each person's body instrument is unique, it is more difficult for singers to monitor and pinpoint their performance faults.

It is in general more difficult for singers to monitor and pinpoint their performance errors since inner body movements cannot be seen directly, and each person's body instrument is quite unique.

Breathing is commonly considered the manifestation of the overall inner body movements during singing as well as the foundation of effective singing techniques[4]. It involves several organs such as the lungs, the diaphragm, and external and internal intercostal muscles, and is closely connected with pitch, timbre, and dynamic control[5][6]. When practicing a song, vocal coaches often carefully listen to the voice and give verbal feedback on breathing states. But even with the vocal tutor's hints and singing demonstration, learners may still get lost on how to control the exact breathing volume and position, let alone when self-practicing. Furthermore, breath control training is shown to improve the singing ability of elementary students[7] and has potential in the singing tutoring interface.

To this end, we design a multimodal singing tutoring interface for elementary singing learners to guide their singing breath. The system mainly consists of 1) a wearable device for singing breathing detection 2) a pitch and breathing analysis model, and 3) a visual interface for singing instruction with real-time pitch and breath feedback. The wearable device uses pressure sensors to detect the movement of ribs, lower abdomen, and back waists to measure the breathing state. The analysis model in the real-time process the singing breath states and fundamental frequencies, comparing them with the score.  The visual interface displays the music score and breath hints at the background, and displays the user's actual breathing states and pitch in the frontend as the feedback, using which one can learn to adjust the vocal performance. Our user study shows that our system can not only adjust learners' breathing states but also improve the average pitch accuracy of musical background learners by 21.25%.

In summary, we contribute

- A haptic interface to effectively detect breathing states during singing in real-time.
- Intuitive visualization of breathing as breath feedback and guidance.
- A user study to quantitatively prove that breathing guidance improves pitch accuracy.

## Related Works

We review three realms of related works: 1) singing tutoring interface, 2) body sensors, and 3) haptic guidance in musical instruments.

### 2.1 Singing Tutoring Interface

A significant number of singing tutoring interfaces focus on the evaluation of vocal qualities which can be detected by computer audio analysis, such as pitch accuracy[8][9], formants[10], harmonics[9], vibrato[11][12], larynx closed quotient[12], and vocal energy[13]. Receiving the above scores help with evaluating the performances, but the learners need more direct tutoring on how to improve their basic singing skills.

## 2.2 Body sensors

There exist some bio-detection devices that facilitate visual interfaces to monitor the users' body states. For example, Katz et al.[14] used several position sensors in the tongue and made real-time 3D animation of tongue movement, and the system was effective in improving speaking articulation. Pettersen et al.[15] examined neck and shoulder muscle activity and thorax movement by strain gauge sensors, and explored the connection between body movement in different pitches and vocal loudness. Cotton et al.[16] captured and used biofeedback signals and physiological changes of singing breathing to generate sound in performance. However, none of the research applies body detection in singing tutoring and explores how the visualization of body movement affects learning.

## 2.3 Haptic Guidance in Musical Instruments

Several works have proposed different haptic devices for supporting music instrument learning. In boosting rhythm accuracy, Holland et al.[2] introduced a wearable device to provide drum learners rhythm hints by vibrotactile stimuli on wrists and ankles, and Tom et all.[1] examined the types of haptic stimuli will have different influences on rhythm learning. Some wearable haptic interfaces are also used in correcting the posture of players, for example, in improving violin novices' straight bowing technique[17] or trumpet player's poor gestures[2]. Learning efficiency can also be increased by using a visual or haptic interface in flute learning[3][18].

## System Design

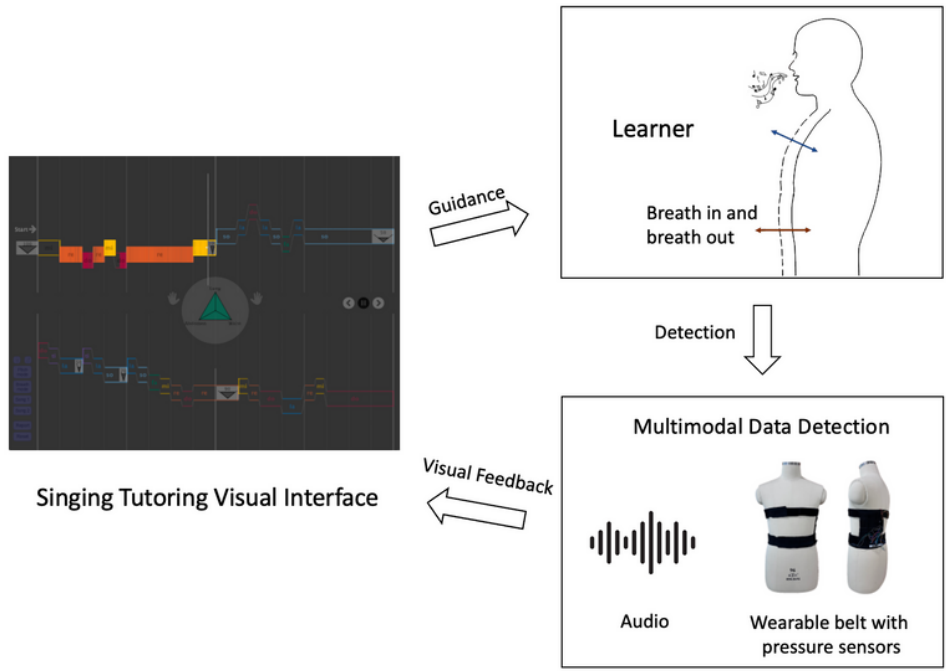

**Figure 1**
The system diagram.

The system diagram is shown in Figure 1. The system displays the music score with tonic sol-fa and breath hints in the background. Based on respiratory mechanics in singing[5], the wearable device uses pressure sensors to detect the movement of ribs, lower abdomen, and back waists to measure the breathing state. Finally, the detected breathing movement data and pitch frequency is processed and visualized to help learners to know their breathing pattern and pitch performance. The rest of this section is organized as follows. We first present the hardware for breath detection in 3.1, then introduce data acquisition and analysis in 3.2, and finally, present the visual interface representation in 3.3.

## 3.1 Hardware Design

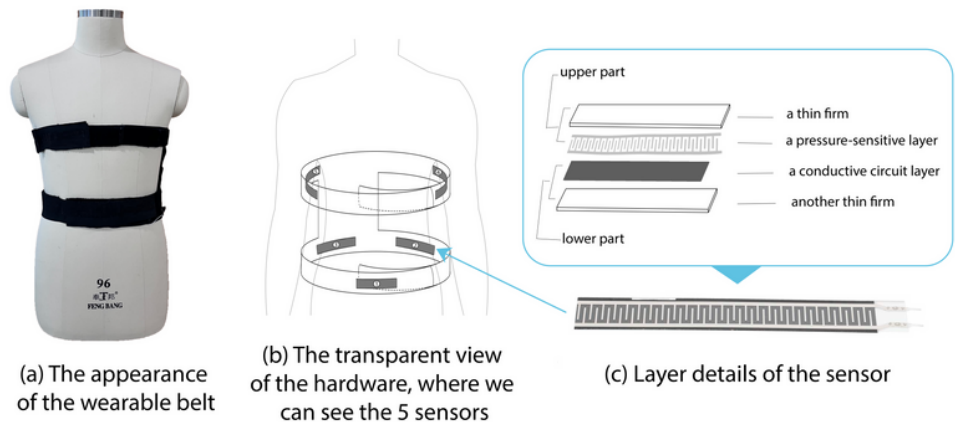

**Figure 2.**
An illustration of the wearable breath detector.

Figure 2 shows our hardware design of the breath detector. The overall design principles are 1) to fit most users' bodies and be comfortable to wear; 2) to accurately detect learners' breathing states. To this end, we chose nylon with elastic rubber bands as the main body of the wearable device (as shown in Figure 2(a)). There are five RP thin-film bend pressure sensors with 110 mm long and five resistance-force conversion modules series connected with the sensors. The pressure sensors are separately positioned on the front of the lower abdomen, two back waists, and two sides of ribs (as Figure 2(b) shows). There are two parts of the pressure sensor: a thin film and a pressure-sensitive layer on the upper part of the sensor, and a thin film and a conductive circuit on the lower part (as Figure 2(c) shows). When outside force applies to the active area, the disconnected circuit of the lower layer will be connected through the pressure-sensitive layer of the upper layer, by which to convert force into resistance.

We set stretchy velcro on the overlapping sides of the belt to adjust the length and ensure a tight fit with the learner's body. To better match each sensors' position to different figures, we place a large area of loops to the inner layer of the belt, so sensors with hooks can be adjusted to the suitable place.

## 3.2 Data Acquisition and Analysis

During inhalation and exhalation, the cross-sectional area of the lower chest and/or the lower abdomen will move, and the sensors will output resistance which can be transferred to force. The minimum responsible force of the sensors is 0.196 N, and it can respond in less than 10 ms. The range of the force exerted in the body is from 5 N to 60 N while breathing. Second, the real-time force data will be transmitted from the Nano BLE microcontroller to our Processing main interface via USB port communication. Then, we calculate the average of the force on the left and right back waists, and the left and two ribs. Considering the differences between body shapes, each frame's sensors data will remove the maximum exhalation status value. After we get the force data to measure the expansion of the lower abdomen (LA), back waist (BW), and ribs (RB), we separately use the mean filter over a window of 150 milliseconds on LA, BW, and RB to remove noises.

As for the pitch estimation, we record 50 milliseconds of real-time singing audio, do Fast Fourier Transform and get the highest energy bin. The naive but fast algorithm works well in detecting monophonic singing voices and can reach over 95% accuracy (tested by a professional singer who sang strictly according to the score).

## 3.3 Visual Interface

Figure 3 shows the preview state and the playing state of the entire visual interface. There are three main parts of the interface: pipe score, breathing circle man, and control panel. A user can choose the piece of song to practice on the left control panel. One can also start to play, pause, or jump to any measure when practicing using the right control panel.

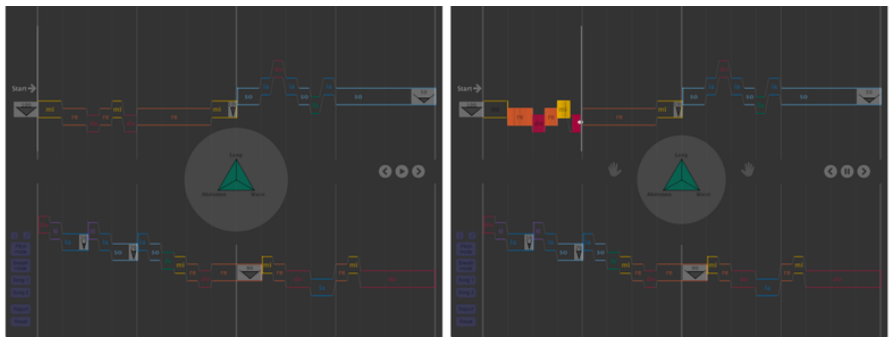

**Figure 3.**
The visual interface of our system.

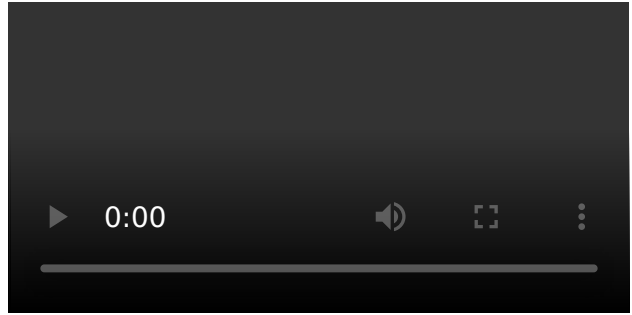

A two measures example of the visual interface

### 3.3.1 Pipe Score

In the background, each page has two rows of hollow Pipe Score, and each row has four measures. Within each measure, there are light vertical lines indicating beats, and a time position bar to follow the track of the song. The Pipe Score marks syllable names of notes and uses note height and rainbow color[18] to visualize different pitches, and note length to indicate rhythm. Each note in a pipe has a small pitch range for trifling voice changes, such as vibration or negligible tune inaccuracy. A closed pipe means a singing sentence, and before each sentence pipe, there is a gray breathing block to hint to the learners how much air they need to inhale. For example, 50 with half "cup" of luftpause means: try to breathe in half a lungful of air.

### 3.3.2 Breathing Circle Man

The Breathing Circle Man has a big circle to represent the belly, and two hands to guide the real-time rhythm. After starting to practice a song, the middle breathing circle will give the inhalation signal by expanding and the exhalation signal by shrinking. Along with the beats metronome sound, the two hands will act like a conductor's hands and tutor the rhythm.

### 3.3.3 Real-time Visual Feedback

On the visual interface, we use a white dot to indicate the player's real-time pitch performance. If the white dot is between the upper and lower edges of a note, then the blank note will be filled in by the note's color, while a mistake note will remain blank.

In the middle of the Breathing Circle Man, there is a green triangle that demonstrates the real-time breathing state. The top vertex of the triangle shows the body movement of ribs and the below two vertices show the body movement of the lower abdomen and back waists.

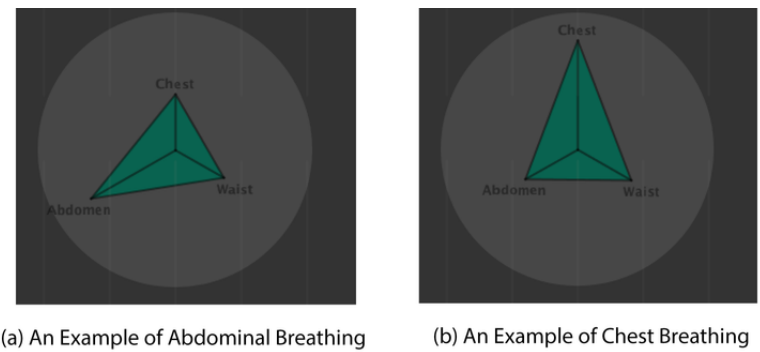

**Figure 4.**
Examples of abdominal and chest breathing patterns.

As figure 4 shows, the best breathing state in singing is a good breathing state, which shows:

1. the length of the abdomen angle bisector will increase when inhaling,
2. the length of the rib angle bisector is basically unchanged

The worse breathing state in singing is a bad breathing state, which shows:

1. the length of the abdomen angle bisector is basically unchanged
2. the length of the rib angle bisector will increase when inhaling

As for the back waist, people may show quite different patterns. Some people may partly get support from the waist but others show no movement in the waist. The shape of the waist could still provide a reference for some people to sense their breath.

## Experiment

We design a comparative experiment to validate the effectiveness of the breath guidance interface. In the experiments, subjects learn to sing in two modes: 1) pitch-only mode, which means there are only Pipe Score guidance and real-time pitch visualization, and 2) pitch+breath mode, which means the interface has both pitch and breath detection and visualization. Then, we conduct a user study that compares the result of pitch-only mode and pitch+breath mode to test the effectiveness of breath detection and visualization in improving breath pattern and pitch accuracy.

### 4.1 Participants

14 subjects (9 males and 5 females with different body shapes) between the ages of 18 and 26 participated in the study. Among the participants, 10 have played instruments in the past (including the piano, the guitar, the violin) but with *no* vocal training experience; 4 have no musical background. All of them are unfamiliar with our interface. Also, all participants reported normal hearing and speaking.

## 4.2 Experiment Design

The experiment employed a 2×2 within-subject factorial design. The independent variables were the Learning mode (pitch-only mode, pitch+breath mode) and Learning piece (Song A, Song B). Each participant learned to sing both song A and song B: one song was learned through pitch-only mode, and the other was learned through pitch+breath mode. In each permutation, there are more than 2 participants with musical backgrounds and 1 participant with no musical background.

## 4.3 Task and Procedure

The procedure consisted of pre-training, learning & testing, and interview.

**Pre-training:** In this step, we first asked all participants to report their musical backgrounds. Then, we introduced to them the procedure of the experiment and the breath-related representation of our visual interface.

**Learning & testing:** After the pre-training step, all participants learned pieces via our tutoring system themselves. The subjects were allowed to listen to the ground truth of scale in piano or demonstrative singing at any time. In pitch+breath mode, they were asked to wear the haptic belt by themselves, took several deep breaths until they were familiar with the device, and then clicked on a button to record sensors values after exhaling all air in their body. Each music piece could be repeatedly practiced until i) the piece is sung accurately, or ii) the practice time reached the maximum limit (15 minutes).

The *pitch accuracy score* is computed as the number of correctly singing notes. A correct note is decided by 1) the pitch is within the range of the correct pitch of a quarter-tone, and 2) the duration of the correct pitch is greater than 10% of the whole note's duration. We saved a data report of the breathing and pitch state of each frame for later analysis of breathing dynamics.

**Interview:** After the practice session, we conducted an interview about the helpfulness of breathing guidance, likes and dislikes compared with vocal coaches and karaoke applications, and further suggestions.

## 4.4 Music To Learn

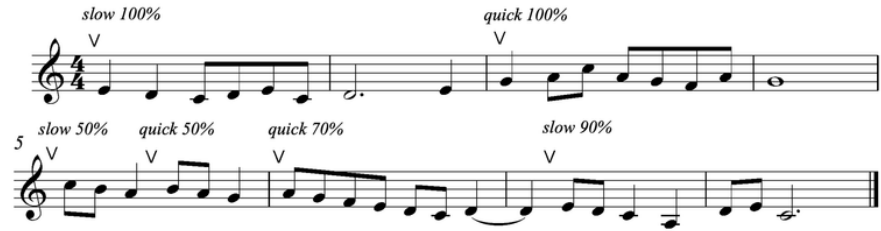

(a) Song A

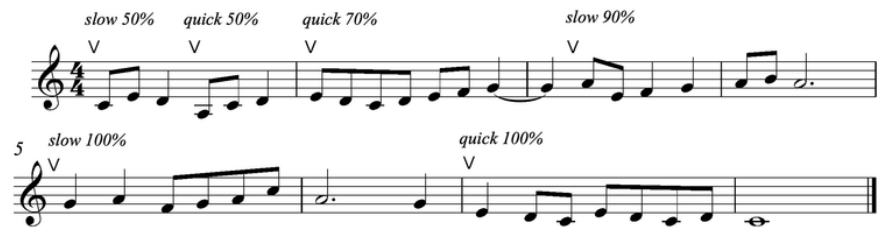

(b) Song B

Figure 5.
The two music pieces with breathing guidance.

We composed two 8-bar pieces tailored to avoid the learning effects in the within-subjects experiment(as shown in figure 6). We ensured that the two pieces are of equal difficulty in the following three aspects:

- The same pitch range (A3 to C5), and cover all pitches in the range
- The same number of inhalation points with the same breathing speed and volume
- The same statistics of intervals as shown in Table 1

**Table 1. Basic statistics of the pitch interval from each note to the next note, approximately reveal the learning difficulty of the two pieces.**

| Pitch Interval (semitone) | Count |
|---|---|
| 1 | 1 |
| 2 | 21 |
| 3 | 5 |
| 4 | 4 |
| 5 | 2 |

Compared with the learning piece used in[19], here we increase the difficulty by adding some jumping notes and more breathing variations.

## 4.5 Result and Discussion

We analyze the overall performances of the 10 participants with musical backgrounds and 4 participants with no musical training in two learning methods.

### 4.5.1 Comparision of Pitch Accuracy Scores

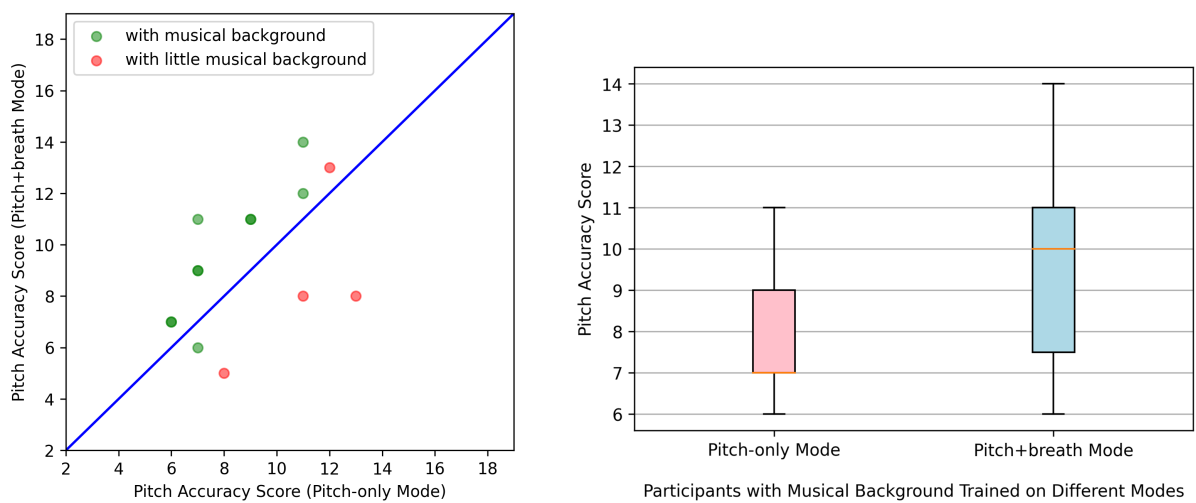

**Figure 6.**
The overall pitch accuracy improvement (with and without breath detection and visualization).

Because we asked all participants to sing in "i" or "u", voice formants have nearly no influence on pitch detection accuracy. We define pitch accuracy as the percent of notes sung correctly. Figure 6 summarizes the pitch accuracy in pitch-only mode and pitch+breath mode. The green points are 10 participants with musical backgrounds, and the 4 red points are participants without musical backgrounds. For musical background participants, it shows that breath visualization and guidance can significantly improve their pitch accuracy in singing, with p = 0.003 < 0.005 by pairwise t-test. All but one point are above the y = x line. The improvement of average pitch accuracy using breath detection and guidance is 21.25% higher than without breath guidance. The large improvement shows that our breathing guidance can help improve pitch accuracy. For participants with no musical background, breathing detection and guidance has no improvement in their pitch accuracy.

### 4.5.2 Comparison of Breathing Patterns

We define Breathing Dynamic Range (BDR) as the difference between the maximum force value and the minimum force value during a piece's practice. We see an enlarged dynamic range as an indication of breathing improvements for beginners. Again, we compare the Breathing Dynamic Range of pitch-only mode and pitch+breath mode in all participants.

The average BDR is 15.38 with a standard deviation of 13.12. Among the 10 participants with musical backgrounds: 7 increased 2.52 Breathing Dynamic Range of the lower abdomen, which shows they start to give more power to the lower abdomen; 6 participants decreased the Breathing Dynamic Range of the ribs expansion by 0.44, which means they reduce the chest breathing portion; 4 participants show both increasing abdomen support and reduced chest breathing portion. However, in the 4 participants with no musical backgrounds, the result shows that their breathing patterns are not improved.

**Table 2. The number of participants who improved their breathing after practicing**

|  | Number of All Participants | Lower Abdomen Support Improved | Ribs Portion Reduced | Both |
| --- | --- | --- | --- | --- |
| Participants with Musical Background | 10 | 7 | 6 | 4 |
| Participants with no Musical Background | 4 | 1 | 1 | 0 |

### 4.5.3 Interview

We conducted interviews with 14 participants and selected some representative remarks that supplement our results.

**Q1: While singing, what do you think of the effect of breathing detection and visualization?**

- "I unconsciously straightened up my body, which can help to make my breathing more open and easier. With breathing guidance, I feel my pitch control is better."
- "This breathing detection is sensitive and the visualization can accurately show my breathing state. I like the breathing triangle representation, and I think this feedback is helpful for practicing singing."
- "I paid more attention to whether my real-time breathing state is abdominal or chest breathing."
- "At first, it's hard for me to focus on pitch and breath at the same time. But after being familiar with the melody, it helped me a lot to breathe better in singing."

**Q2: Compared with singing coaches and some commercial karaoke smartphone apps, is there any aspect in our system that you like or think needs to be improved?**

- "It is abstract to sense the position of abdominal breathing from the singing coach's guidance. Compared with that, the visual interface is interesting and can directly show whether my breathing is from the chest or abdomen."
- "I felt a little bit nervous when making some mistakes with the tutoring system. I suggest adding some encouraging sentences so I can keep practicing."
- "I don't like the rating of some commercial applications, which may lead you to care more about the score. This singing guidance system purely focuses on practicing and improving singing skills. I would like to long term use it to improve my breath."

**Q3: Any other suggestions?**

- "Now the score gives fixed inhalation points and volume. It would be better to provide different instructions on different people's breathing state in singing."
- "It will be more helpful for me to give a detailed report about my breathing mistakes or suggestions."
- "The song is difficult for me. It would be slower or have different levels of songs to learn"
- "After singers can inhale to the abdomen, it would be better to guide them on how to exhale correctly and evenly."

## Conclusion and Future Work

We have contributed a multimodal singing tutoring system featuring breath guidance. The system guides breathing in real-time by first detecting the breath using a newly designed wearable device and then visualizing the breathing states on an interactive score interface. Our experiment shows that our system can help people improve their diaphragmatic breathing in singing especially for participants with musical backgrounds. The participants with no musical background can hardly sing the correct pitch, let alone notice the breathing visualization and guidance. The overall pitch accuracy is 21.25% higher than without breathing detection and visualization.

According to our interview, we would like to improve our breath guidance interface in the following aspects in the future. First, we will modify the wearable belt by combining different kinds of haptic feedback. Second, we will improve the visual interface by providing more detailed and tailored breathing suggestions. Last but not least, we see this study as a good starting point to explore the science of singing breath in a human-computer paradigm, discovering the relationship between breathing, music, and the human body.

## Acknowledgments

We want to thank all the participants who joined our study and the support of Music X Lab. We would also like to thank Yinmiao Li and Daniel Chin for their kindly help and suggestions.

## Citations

1. Tom, A., Singh, A., Daigle, M., Marandola, F., & Wanderley, M. (2020). Haptic Tutor-A haptics-based music education tool for beginners. *International Workshop on Haptic and Audio Interaction Design*. ↵
2. Dalgleish, M., & Spencer, S. (2014). Postrum: developing good posture in trumpet players through directional haptic feedback. *The 9th Conference on Interdisciplinary Musicology*. ↵
3. Xia, G., Jacobsen, C., Chen, Q., Yang, X., & Dannenberg, R. (2018). ShIFT: A Semi-haptic Interface for Flute Tutoring. *International Workshop on Haptic and Audio Interaction Design*. ↵
4. Leanderson, R., & Sundberg, J. (1988). Breathing for singing. *Journal of Voice*, *2*(1), 2–12. ↵

5. Sundberg, J. (1992). Breathing behavior during singing. *Stl-Qpsr*, *33*, 49–64. ↵
6. da Costa, T. D., Vara, M. de F. F., Cristino, C. S., Zanella, T. Z., Neto, G. N. N., & Nohama, P. (2019). Breathing monitoring and pattern recognition with wearable sensors. In *Wearable Devices-the Big Wave of Innovation*. IntechOpen. ↵
7. Phillips, K. H. (1985). The Effects of Group Breath-Control Training on the Singing Ability of Elementary Students. *Journal of Research in Music Education*, *33*(3), 179–191. https://doi.org/10.2307/3344805 ↵
8. Lin, K. W. E., Anderson, H., Hamzeen, M., & Lui, S. (2014). Implementation and evaluation of real-time interactive user interface design in self-learning singing pitch training apps. *International Computer Music Conference*. ↵
9. Mayor, O., Bonada, J., & Loscos, A. (2009). Performance analysis and scoring of the singing voice. *Proc. 35th AES Intl. Conf., London, UK*, 1–7. ↵
10. Angelakis, E., Kosteletos, G., Andreopoulou, A., & Georgaki, A. (2018). Development and Evaluation of an Audio Signal Processing Educational Tool to Support Somatosensory Singing Control. *Audio Engineering Society Convention 145*. Audio Engineering Society. ↵
11. Mayor, O., Bonada, J., & Loscos, A. (2009). Performance analysis and scoring of the singing voice. *Proc. 35th AES Intl. Conf., London, UK*, 1–7. ↵
12. Ventura, J., Sousa, R., & Ferreira, A. (2012). Accurate analysis and visual feedback of vibrato in singing. *2012 5th International Symposium on Communications, Control and Signal Processing*, 1–6. IEEE. ↵
13. Rossiter, D., & Howard, D. M. (1996). Albert: a real-time visual feedback computer tool for professional vocal development. *Journal of Voice: Official Journal of the Voice Foundation*, *10*(4), 321–336. ↵
14. Rossiter, D., & Howard, D. M. (1996). Albert: a real-time visual feedback computer tool for professional vocal development. *Journal of Voice: Official Journal of the Voice Foundation*, *10*(4), 321–336. ↵
15. Bell, D. (1996). Using Digital technology in a voice lesson. *Canadian Acoustics*, *24*, 34–34. ↵

16. Katz, W., Campbell, T. F., Wang, J., Farrar, E., Eubanks, J. C., Balasubramanian, A., … Rennaker, R. (2014). Opti-speech: A real-time, 3D visual feedback system for speech training. *Fifteenth Annual Conference of the International Speech Communication Association*. Citeseer. ↩

17. Pettersen, V., Bjørkøy, K., Torp, H., & Westgaard, R. H. (2005). Neck and shoulder muscle activity and thorax movement in singing and speaking tasks with variation in vocal loudness and pitch. *Journal of Voice*, *19*(4), 623–634. ↩

18. Cotton, K., Sanches, P., Tsaknaki, V., & Karpashevich, P. (2021). The Body Electric: A NIME designed through and with the somatic experience of singing. *International Conference on New Interfaces for Musical Expression*. PubPub. ↩

19. Dalgleish, M., & Spencer, S. (2014). Postrum: developing good posture in trumpet players through directional haptic feedback. *The 9th Conference on Interdisciplinary Musicology*. ↩

20. Tom, A., Singh, A., Daigle, M., Marandola, F., & Wanderley, M. (2020). Haptic Tutor-A haptics-based music education tool for beginners. *International Workshop on Haptic and Audio Interaction Design*. ↩

21. Van Der Linden, J., Schoonderwaldt, E., Bird, J., & Johnson, R. (2010). Musicjacket—combining motion capture and vibrotactile feedback to teach violin bowing. *IEEE Transactions on Instrumentation and Measurement*, *60*(1), 104–113. ↩

22. Dalgleish, M., & Spencer, S. (2014). Postrum: developing good posture in trumpet players through directional haptic feedback. *The 9th Conference on Interdisciplinary Musicology*. ↩

23. Xia, G., Jacobsen, C., Chen, Q., Yang, X., & Dannenberg, R. (2018). ShIFT: A Semi-haptic Interface for Flute Tutoring. *International Workshop on Haptic and Audio Interaction Design*. ↩

24. Chin, D., Zhang, Y., Zhang, T., Zhao, J., & Xia, G. G. (2020). The Body Electric: A NIME designed through and with the somatic experience of singing. *International Conference on New Interfaces for Musical Expression*. PubPub. ↩

25. Sundberg, J. (1992). Breathing behavior during singing. *Stl-Qpsr*, *33*, 49–64. ↩

26. Chin, D., Zhang, Y., Zhang, T., Zhao, J., & Xia, G. G. (2020). The Body Electric: A NIME designed through and with the somatic experience of singing. *International

*Conference on New Interfaces for Musical Expression*. PubPub.↩

27. Li, Y., Piao, Z., & Xia, G. (2021). A Wearable Haptic Interface for Breath Guidance in Vocal Training. *International Conference on New Interfaces for Musical Expression*. PubPub. ↩